\documentclass[12pt,a4paper]{article}
\usepackage{graphicx}
\usepackage{amsmath}
\usepackage{graphicx}
\usepackage{amsfonts}
\usepackage{amssymb}

\begin{document}
\author{S. Brazovskii$^{a,\ast}$, and N. Kirova$^{b}$ \\
	$^{1}$Universit\'{e} Paris-Saclay, CNRS, LPTMS, 91405, Orsay, France\\
	$^{2}$Universit\'{e} Paris-Saclay, CNRS, LPS, 91405, Orsay, France\\
	$^{\ast}$e-mail: serguei.brazovski@universite-paris-saclay.fr}
\title{Phase slips, dislocations, half-integer vortices, two-fluid hydrodynamics and the chiral anomaly in charge and spin density waves.}
\date{\it{Contribution for the JETP special issue in honor of I.E.Dzyaloshinskii’s 90th birthday}}

\maketitle

\textbf{Abstract.} This brief review recalls some chapters in theory of sliding incommensurate density waves  which may have appeared after inspirations from studies of I.E Dzyaloshinskii and collaborations with him. First we address the  spin density waves which rich order parameter allows for an unusual object of a complex topological nature: a half-integer
dislocation combined with a semi-vortex of the staggered magnetization. It
becomes energetically preferable with respect to an ordinary dislocation due
to the high Coulomb energy at low concentration of carriers. Generation of
these objects should form a sequence of $\pi$ - phase slips in accordance
with experimental doubling of phase-slips rate.

Next, we revise the commonly
employed TDGL approach which is shown to suffer from a violation of the charge
conservation law resulting in nonphysical generation of particles which is
particularly pronounced for electronic vortices in the course of their nucleation or motion. The
suggested consistent theory exploits the chiral transformations taking into
account the principle contribution of the fermionic chiral anomaly to the
effective action. The derived equations clarify partitions of charges,
currents and rigidity among subsystems of the condensate and normal carriers
and the gluing electric field. Being non-analitical with respect to the order parameter, contrarily the conventional TDGL type, the resulting equations still allow for a numerical modeling of transient
  processes related to space- and spatiotemporal vorticity in DWs.

\section{ Introduction.}

\subsection{Inspirations from I.E. Dzyaloshinskii.}

The authors had a chance to publish together with I.E. Dzyaloshinskii
(IED in the following) the article \cite{SB:81} on doubly-quasi-periodic
solitonic lattices emerging in a 1D electronic system under simultaneous
effects of the charge doping away from the half band filling and of the spin
polarization. This publication was in the course of our studies in theory of
charge density waves (CDW) which had been started in 1976 by one of us (SB, a
thankful disciple of IED) under the inspiration and initially with
participation \cite{SB:76,SB:77} of IED. The trick of the chiral
transformations having been emploied in \cite{SB:76} (see below in Sec.3.2), actually suggested
by IED, provided a handy frame to study adiabatic models of CDWs which later would
lead SB to notice an instability of normal electrons or holes, excited or
injected to a CDW, towards formation of solitons with the electron buried at
the solitons' midgap state, see \cite{BK:84} for a review. The resulting physics
of microscopic solitons and their arrays \cite{SB:09} can be traced back to
IED invention of the commensurability locking (published neglectfully only in
conference proceedings \cite{IED:73}) and forth to his work on exact solutions
of many-body lattice models
\cite{IED+SB+IK:82-jetp,IED+SB+IK:82,IED+IK:82,IED+IK:83}. Much later, an
indirect inspiration came to us \cite{KB:99-sdw,KB:00-sdw} from the article
\cite{IED:77} where IED had noticed that a presence of a dislocation in a
crystal possessing the antiferromagnetic spin order enforces a curious
half-integer vortex of spins' rotations.

In spite of an antiquity and following periods of intensive developments,
these old studies have left some intriguing openings and unresolved problems
which we succeeded to formulate only later in \cite{KB:99-sdw,KB:00-sdw,BK:99-cdw} and to detail very recently
in \cite{SB+NK:19-annals,SB+NK:19-jetp}. In this short review we shall
address some of these issues. First, we shall describe combined topological
defects - half-integer vortices of displacements and spin rotations in
incommensurate SDWs which might necessarily appear under applications of the driving
electric field. These considerations are related to IED work \cite{IED:77} and
are coherent with a persistent interest in fractional vortices: from Helium A
(\cite{Volovik:76,Cross:77} about the same time as the IED work) and
triplet superconductors (reviews \cite{Luk:95,Volovik:00}) to FFLO
phase \cite{Radzihovsky:09,SB:09-FFLO} and Bose-condensate of
polaritons \cite{Rubo:07}, see also \cite{Mineev:13} for a more recent review
of the literature taking into account also experimental attempts.

Our results on SDWs have been published only briefly in conference
proceedings \cite{KB:99-sdw,KB:00-sdw}; here we shall update them,
particularly adding new results on a numerical modelling of static
vortices-dislocations and nonstationary phase-slip processes (1+1 space-time
vortices). Next we shall quote and augment for modelling quite a recent
development \cite{SB+NK:19-annals} on construction of the chirally invariant
(recall \cite{SB:76}) description of transient processes in a CDW or a SDW in presence
of normal carriers. This formulation is free from a drawback of the non-conservation of condensed particles which we show to be inherent to the traditional TDGL approach. Beyond the frame of this article, recall also the most recent work
\cite{SB+NK:19-jetp} (devoted to ours top teacher - I.M. Khalatnikov)  on a hydrodynamics of a DW at presence of an ensemble of
intrinsic defects - phase solitons or dislocation loops. This direction is in
line with long standing interests of IED in topological defects, recall \cite{IED:70-LC} and particularly \cite{IED+GV}.

\subsection{Significance of static and transient topological configurations in incommensurate sliding density waves.}

Embedded or transient topologically nontrivial configurations are common among
symmetry broken ground states of electronic systems. Among them, beyond
vortices and the FFLO state in superconductors, there is a vast family of
electronic crystals, including charge density waves (CDWs),
spin density waves (SDWs), Wigner crystals and stripes arrays; with CDWs, and
also SDWs, presenting the most popular and convenient object of studies (see
reviews \cite{Monceau:12,Gruner-CDW,Gruner-SDW,Gorkov+Gruner}).

A unique property of electronic crystals is related to their ability for the
collective current conduction by means  of  sliding. It is ultimately related to
appearance, under stresses from the electric field or imbalances of normal
particles, of topologically nontrivial objects like solitons, vortices,
dislocations, and transient processes of phase slips; all that gives rise to a
rich complex of nonlinear and nonstationary behavior. We refer to recent
publications \cite{SB+NK:19-jetp,Karpov:19} for a literature review
concerning modern experimental manifestations and various theoretical contributions.

The ground state degeneracy allows for configurations connecting equivalent
while different states across a disturbed area. These configurations are
protected either topologically or by conservation laws of the charge or the
spin. They are known commonly under a generic name of "topological defects"
which includes extended objects like planes of domain walls as arrays of
solitons \cite{BK:84,SB:07}, lines or loops of dislocations
\cite{Feinberg:88,SB+SM-disl:92,Hayashi:96} as phase vortices, and local
objects like phase and amplitude solitons \cite{BK:84,SB:09,SB:90}.

The density waves (DWs) are seen as superstructures, usually weak $A\ll1$ and
hence harmonic $\sim A\cos[\vec{q}_{0}\vec{r}+\varphi(\vec{r})]$, produced by
modulations of electronic charges and atomic displacements or of electronic
spins, for CDWs or SDWs correspondingly. In most common quasi-1D
systems the density wave (DW) is a kind of an elastic uniaxial crystal, which
displacements are given by distortions of the DW phase $\varphi(\vec{r})$.
Actually, phase displacements can be unlimitedly large which brings about such
spectacular phenomena as the gigantic permittivity (in this respect the DW can
be interpreted as a ferroelectric sitting always exactly at the transition
temperature) and even more striking effect of the collective Froehlich
conductivity due to the overall sliding with the collective current
proportional to the phase velocity $j_{c}\propto\partial_{t}\varphi$. The
DW sliding is ultimately related to the current conversion process which
necessarily involves topological defects like solitons, phase slips,
dislocation lines/loops. The conversion of the normal current to the
collective one passes through consecutive dynamic and kinetic steps:
free electrons injected from the contact transform dynamically to amplitude
solitons (see for review \cite{BK:84,SB:90}) which keep
carrying the electron's spin 1/2 while the local charge is compensated to zero. Pairs of amplitude solitons merge into spinless $2\pi$ phase solitons
of the even lower energy, thus forming local microscopic phase slips. Their
subsequent aggregation forms edge dislocation lines (D-lines) or loops
(D-loops) expanding across the sample, thus completing the conversion of
excess concentration of normal carriers to new periods of the DW
superstructure. Macroscopically, the phase slip develops as the edge D-line
proliferating/expanding across the sample \cite{Ong:84,Ong:85} or as the plane
vanishing of the DW amplitude across a narrow sample
\cite{Gorkov:83,Gorkov:89,Gorkov:84,Batistic:84}. Beyond phase
slips, the dislocations are supposed to participate in depinning via creation
of metastable deformations in the course of the DW sliding over the host
lattice imperfections \cite{Larkin:95,SB+TN:04}, in the "Narrow Band
Noise"  generation, and also arise in contact structures. The point defects
($\pm2\pi$-solitons as "add-atoms" and "vacancies" or also as nucleus dislocation
loops) compete with electrons as normal carriers, also providing a material
for the climb of dislocations required for their expansion.

Properties of dislocations are strongly affected by Coulomb forces and hence depend on
screening facilities of free carriers. The DW deformations generate a local
charge density $n_{c}\propto\partial_{x}\varphi$ which brings about a high
cost of the Coulomb energy. The Coulomb enhancement of the dislocation energy
plays an intriguing role in Spin Density Waves bringing to life a kind of a
mixed topological object: a half - integer dislocation combined with a semi -
vortex of a staggered magnetization vector. The phase slips, necessary for the
current conversion or the depinning, should proceed via propulsion of these
combined objects which provides a natural interpretation for a confusing
2-fold enhancement of a frequency generated by moving Spin Density Waves in
comparison with Charge Density Waves.

Beyond transient dislocations contributing to phase slips, static arrays of
dislocations can appear in lateral geometries when the electric field is
applied transversely to the direction $x$ of the CDW sliding. Indeed, the
interaction energy $e\Phi\partial_{x}\varphi/\pi$ of the deformed CDW with the
electric potential $\Phi$ resembles the one for a superconductor under a
magnetic field described by the vector potential with the component
$A_{x}\equiv\Phi$. Then the transverse, to the chains, electric field
$E_{y}=-\partial_{y}\Phi$ acts upon the CDW phase analogousely to the action
of the transverse magnetic field in a superconducting film. Hence an array of static
dislocations should appear above the electric field threshold
as the vortex lattice above $H_{c1}$ in a superconductor. Nevertheless the situation in a CDW is more complicated
technically because of a strong intrinsic field generated by D-lines which
requires for a self-consistent treatment contrary to the common practice in superconductors were the
magnetic field is commonly treated as the external one. The physics of vortices
in CDWs is also more complex because of the interaction of the CDW phase also
with the concentrtion $n$ of normal carriers via the energy $n\partial_{x}\varphi\hbar v_{F}/2$
describing the breathing of the Fermi level following the CDW deformations.

Figures 1 and 2 illustrate the sequences of phase slips and the
 static dislocation. The plots were obtained from numerical modelling
of a TDGL type equations as it has been described elsewere
\cite{SB+NK:19-annals,Yi:13,Yi:15}.

\begin{figure}[tbh]
	\includegraphics{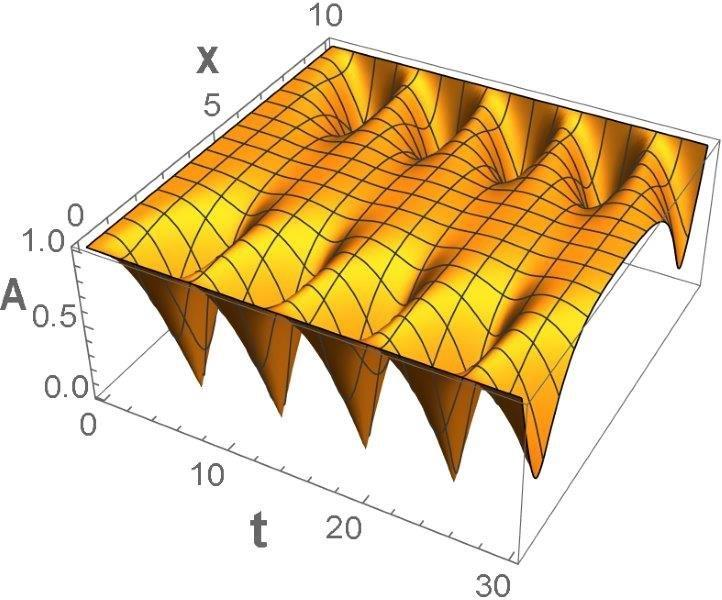}
	\caption{ A static array of dislocations appearing in the electric field applied transversely to the direction $x$ of DW displacements. }
	\label{fig:1}
\end{figure}

\begin{figure}[tbh]
	\includegraphics{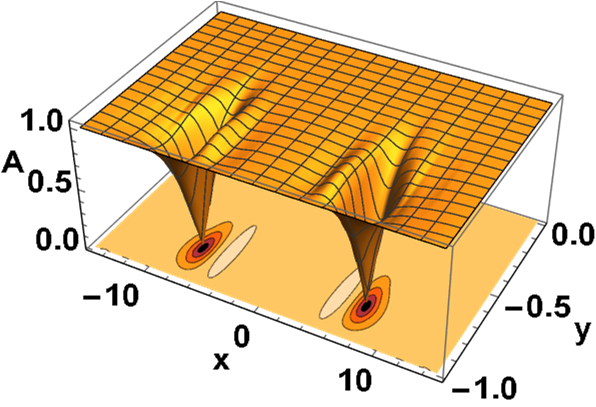}
	\caption{ A static array of dislocations appearing in the electric field applied transversely to the direction $x$ of DW displacements. }
	\label{fig:2}
\end{figure}

\section{ Combined half-integer dislocation and the magnetization
vortex in a spin density wave.}

CDW and SDW are characterized by scalar and vector order parameters:
$\eta_{cdw}(\vec{R})=A\exp(i\varphi)$ and $\vec{\eta}_{sdw}=A\vec{m}\exp(i\varphi)$,
where $\vec{m}$ is the unit vector of the staggered magnetization and $\vec{R}=(x,\vec{r}_{\perp})$ with $x$ being the chain direction,. Here we
will show that SDWs allow for, and actually prefer, an unusual $\pi$ phase
vortex which is forbidden in CDWs where only $2\pi$ vortices are allowed
\cite{KB:99-sdw,KB:00-sdw}. Namely, in SDW conventional dislocations loose
their priority in favor of complex topological objects: a half-integer
dislocation combined with a semi-vortex of the staggered magnetization vector, as illustrated in Figure 3.

\begin{figure}[tbh]
	\centering
	\includegraphics{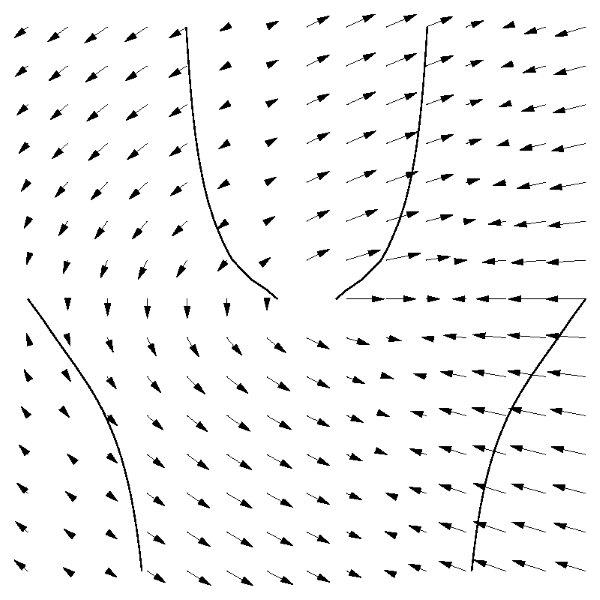}
	\caption{Vector-field $\vec{m}$ for the half-integer dislocation combined with the semi-vortex of the magnetization. Solid lines indicate constant DW phases. Due to the presence of the half-integer dislocation, the number of sites changes between the upper and the lower rows from 2.5 periods (6+6+3) to 2 periods (7+5).}
	\label{fig:3}
\end{figure}

A possible manifestation of half-integer vortices can be found in the NBN generation originated by a periodical sequence of phase slips in the course of sliding. The $\pi$- phase slip reduces twice the universal ratio $\Omega/j$ of the fundamental NBN frequency $\Omega$
to the mean sliding current $j$ as we shall describe in details below. The
splitting of the normal $2\pi$ -dislocation to the $\pi$ ones is energetically
favorable due to Coulomb interactions which harden the elasticity with
respect to the phase displacements while leaving intact the rigidity with
respect to spin rotations. The magnetic anisotropy confines half-integer vortices in pairs connected by a
magnetic domain wall which binding can be overcome by the Coulomb repulsion.
The energy density of spin rotations $\vec{m}(\vec{R})$ in SDWs is not affected by Coulomb forces:
\begin{equation}
W_{spin}\{\vec{m}\}=\frac{1}{2N_{F}}
\left[\tilde {C}_{\parallel}\left(\partial_{x}\vec{m}\right)^{2}+
\tilde{C}_{\bot}\left(  (\partial_{y}\vec{m})^{2}+(\partial_{z}\vec{m})^{2}\right)
+W_{sa}\right]
\label{W-spin}
\end{equation}
Here $\tilde{C}_{\parallel}$,$\tilde{C}_{\bot}$ are the elastic moduli related
to the rotation of the staggered magnetization unit vector $\vec{m},$ $W_{sa}$
is the spin anisotropy energy. $\tilde{C}_{\parallel}$,$\tilde{C}_{\bot}$ are
similar to bare moduli of phase displacements taken without Coulomb
interactions:  $C_{\parallel}^{0}$,$C_{\bot}$ below in Eq.  (\ref{W-charge}).

The  energy for deformations related to phase displacements in both CDW and SDW takes a form
\begin{equation}
W_{chrg}\{\varphi\}=\frac{1}{2N_{F}}
\left[C_{\parallel}^{0}\left(\partial_{x}\varphi/\pi\right)^{2}+
C_{\bot}(\nabla_{\bot}\varphi/\pi)^{2}\right]  +W_{C}~+W_{str},
\label{W-charge}
\end{equation}
where  $N_{F}=2/(\pi\hbar v_{F})$ and $v_{F}$ are the
density of states and the Fermi velocity of the parent metal. Dimensionless
parameters $C_{\Vert}^{0}$ and $C_{\bot}$ are the normalized compression and
share moduli; $C_{\bot}\sim(AT_{c}N_{F}a_{\perp})^{2}$ is a measure of the
interchain coupling related to the transition temperature $T_{c}$. $W_{str}$
is the stress energy from an applied electric potential and/or from disbalance
of normal carriers which promote deformation and/or the motion of the DW, see
more in Sec.3.

The Coulomb part of the energy $W_{C}$ comes from the local charge density
related to the DW displacements: $n_{c}=e\rho_{c}\partial_{x}\varphi/\pi s$,
where $\rho_{c}$ and $\rho_{n}=1-\rho_{c}$ are the normalized densities of the
condensate and of the normal carriers. The Coulomb interactions
drastically affect the charged phase deformations of dislocations greatly
increasing their energy and stretching the shape in the chains' $x$ direction
\cite{SB+SM-disl:92} as we shall remind below with some more details in the
Appendix. Vaguely, the combined effect of Coulomb interactions and the screening results in
hardening of the effective compessibility \cite{Artemenko:86} which vanishes
at the transition and diverges at low $T$ with freezing out of normal
carriers. The effective compressibility $C_{\parallel}$ hardens with growing
$r_{\perp}$ (starting from $C_{\parallel}^{0}$ at shortest interchain
distances) as $C_{\parallel}\sim r^{2}/r_{0}^{2}$ beyond the screening length
of the parent metal $r_{\perp}>r_{0}\sim1\mathring{A}$, until it saturates
beyond the screening length $r_{scr}=r_{0}/\sqrt{\rho_{n}}$ at the value which
grows activationally with lowering $T$. At $r_{\perp}>r_{scr}$
\[
C_{\parallel}^{0}\Rightarrow C_{\parallel}=\rho_{c}/\rho_{n}\mathrm{:}
~C_{\parallel}\propto\rho_{c}~\propto A^{2}\propto(T_{c}-T)~\mathrm{at}~T\rightarrow T_{c}\]
and
\[C_{\parallel}\propto\rho_{n}^{-1}\propto\exp(\Delta/T)~\mathrm{at}~T\rightarrow 0
\]

The resulting big energy of dislocations at small $\rho_{n}$ does not prevent
their creation which just requires for bigger applied potentials, but the
Coulomb energy changes drastically the energy dependence of the dislocation on
its position $Y$ (with respect to a counterpart or a surface). While the
spin-vortex energy per its unit length is a standard $W_{V}\sim T_{c}\ln(Y/a_{\perp})$,
for the dislocation as the charged phase vortex this form of
the energy is reached only when the Coulomb interaction is screened at $Y\gg r_{scr}$ where
$W_{D}\sim T_{c}(r_{scr}/r_{0})\ln(Y/r_{scr})$ with the energy scale being
greatly enhanced as $r_{scr}/r_{0}$.

Coulomb interactions become even more important in an intermediate (wide at low $T$) region
$r_{0}<r_{\perp}<r_{scr}$ governed by the nonlocal, due to unsceened Coulomb interactions,
elasticity with $C_{\parallel}\sim r^{2}/r_{0}^{2}$. For dislocations with
$r_{scr}\gg Y\gg r_{0}$ a curious confinement law is established with
$W_{D}\sim T_{c}Y/r_{0}$ meaning a constant force acting upon the D-line.

The energies $W_{V}(Y)$ and $W_{D}(Y)$ are similar only near $T_{c}$ at small
$A$, hence the vanishing gap $\Delta$, when the abundant free carriers screen
Coulomb interactions already at shortest distances. Otherwise, at the developed gap $\Delta>T$,
the energy of dislocations is greatly enhances with respect to that of
vortices which brings about the option of their spitting into combined
half-integer vortices. In SDWs the Coulomb enhancement of the dislocation
energy plays a principal role bringing to life a special combined topological
object: the half-integer dislocation $\varphi\rightarrow\varphi+\pi$
accompanied by the $180^{o}$ rotation $\mathcal{O}_{\pi}$ of the staggered
magnetization $\vec{m}\rightarrow-\vec{m}$. Indeed, the SDW order parameter
$\vec{\eta}=\vec{m}\cos(Qx+\varphi)$ allows for the following three types of
self-mapping $\vec{\eta}\rightarrow\vec{\eta}$ associated with vorticities
$\nu_{\phi}$ and $\nu_{m}$.\newline i. normal dislocation
$D_{2\pi}$:
$\varphi\rightarrow\varphi+2\pi\equiv\varphi$ and $\ \vec{m}\rightarrow\vec{m}$,
$\nu_{\phi}=1$ and $\nu_{m}=0$;
\newline
ii. normal $\vec{m}$ - vortex $V_{2\pi}$:
$\vec{m}\rightarrow\mathcal{O}_{2\pi}\vec{m}\equiv\vec{m}$ and
$\varphi\rightarrow\varphi$, $\nu_{\phi}=0$ and $\nu_{m}=1$;
\newline
iii. combined object $D_{\pi}V_{\pi}$: $\varphi\rightarrow\varphi\pm\pi$ and
$\vec{m}\rightarrow\mathcal{O}_{\pi}\vec{m}=-\vec{m}$, $\nu_{\phi}=\pm1/2$ and $\nu_{m}=\pm1/2$.

In the last case both the orientational factor $\vec{m}$ and the translational one
$\cos(Qx+\varphi)$ change the sign, but their product in $\vec{\eta}$ stays invariant, as it is demonstrated schematically in Figure 3. Remind that at a given position the energies of vortices depend on their winding numbers as $\propto\nu_{\varphi,m}^2$

We must compare the energies of objects i. and iii. under the requirement of the charge
conservation i.e. preserving the total phase vorticity $\sum\nu_{\phi}$. For
magnetic vortices the total vorticity $\sum\nu_{m}$ does not need to be
conserved but in the bulk it must be kept zero; otherwise the energy of
noncompensated vortices diverges logarithmically at large distances. For phase
dislocations the energy divergence is not a limitation since it is compensated
by driving potentials. Hence, the only decomposition path for the conventional
dislocation of the case i. to the two pairs of half-vortices of the case iii.
is $D_{2\pi}\Rightarrow\{D_{\pi},V_{\pi}\}+\{D_{\pi},V_{-\pi}\}$. If the
energy parameters for both $D$ and $V$ were the same, then the dissociation
cost is zero: $\nu_{2\pi}^{2}=1\Rightarrow4\nu_{\pi}^{2}=1$ and the result
will depend on a tiny balance of similar coefficients. But with dominating
Coulomb energy of dislocations, as expected at low $T$, the magnetic vortex
energy can be neglected, then the decomposition gains nearly half of the
energy ($\nu_{2\pi}=1)^{2}\Rightarrow2(\nu_{\pi}=1/2)^{2}=1/2$ which makes it
inevitable.

Figures 4,a,b  demonstrate results of our modeling of the evolution from a single phase-only integer vortex - the dislocation - to the split pair of half-integer combined vortices. Figure 4a presents the 3D plot for the SDW order parameter amplitude; Figure 4b presents vector fields of the CDW phase and of the spin rotation on the background of the density plot for the amplitude.

Figure 4.
\begin{figure}[tbh]
	\centering	\includegraphics[width=5cm]{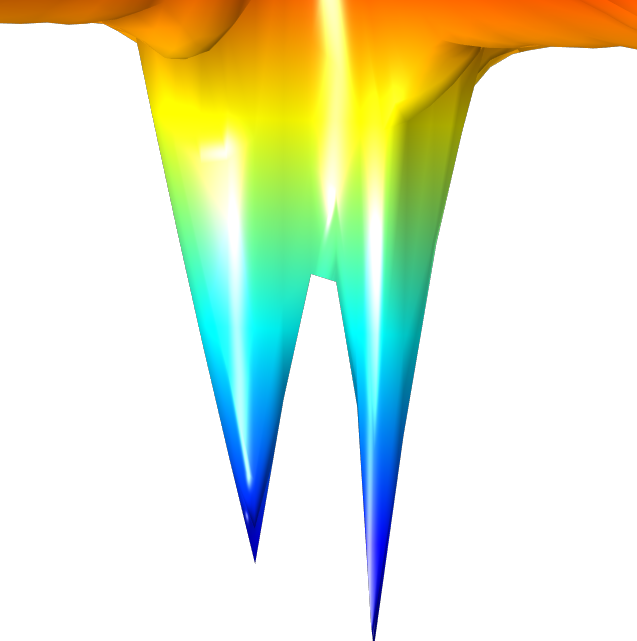}
	\includegraphics[width=5cm]{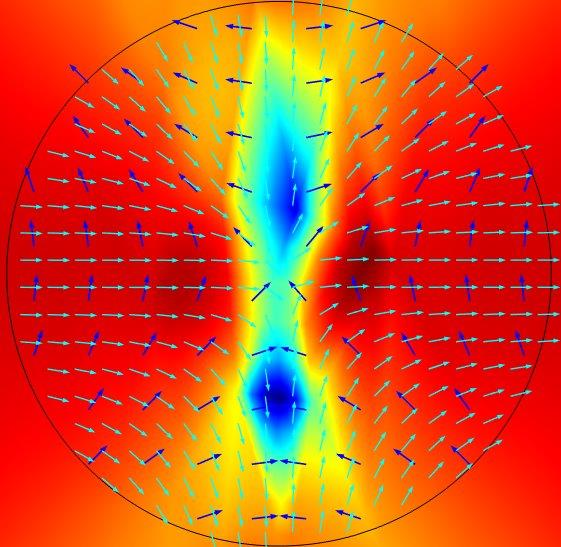}
		\caption{Results of the modeling of the evolution from a single integer dislocation to the split pair of half-integer combined vortices:  the order parameter amplitude $A(x,y)$ (left panel); vector fields of the CDW phase and of the spin rotation on the background of the density plot for the amplitude (right panel).}
	\label{fig:4}
\end{figure}

The spin anisotropy of an "easy plane" type
allows for a free rotation of spins, so it will not affect the above
conclusions. The same will hold in the pure "easy axis" case at presence of a
magnetic field $H>H_{sf}$ exceeding the spin-flop field $H_{sf}\sim1T$ above
which the spins will be tilted, thus possessing a free rotation at the hard
plane. The known SDW crystals possess low symmetries which originates the spin
anisotropy in all three directions. Being small, the anisotropy will not
affect the arrangement in a vicinity of the DL but at large distances from the
DL the $\pi$-rotation of spins will be concentrated in space within the
Ne\'{e}l domain wall, as it has been outlined already in \cite{IED:77}. The
wall will form a string (a plane in 3D) which tempts to confine the two
combined objects.

Unusually, there will be no confinement within a quite wide region $r<r_{scr}$
where the linear law, rather than conventional logarithmic one, for the D-line
energy takes place as we have sketched above and in the Appendix. Here the
total energy gain with respect to the normal DL is $-E_{C}N/2$ where
$E_{C}\sim T_{c}a_{\perp}/r_{0}$ and $N=Y/a_{\perp}$ is the number of chains
separating combined vortices. This repulsive anti-confinement energy directly
overruns the energy lost due to domain wall formation $W_{spin}^{A}=w^{A}N$,
both having the similar $N$ dependence. Usually $w^{A}\sim1K/chain\ll
E_{C}\sim10^{1}K$, so the net interaction between the two semivortices is
strongly repulsive.

Beyond the screening length $r>r_{scr}$, the Coulomb energy slows down while
the $W_{spin}^{A}$ keeps growing linearly, then the total energy gain of two
objects with respect to one DL is
\[
W=-E_0\ln N+w^{A}N~,~E_{0}\sim T_c(r_{scr}/r_0)=T_c/\sqrt{\rho_n}
\]
Hence there is an equilibrium distance between the semi-vortices $N_{eq}\sim
E_{0}/w^{A}\propto1/\sqrt{\rho_{n}})$ which diverges with freezing out of
the screening when $\rho_{n}\rightarrow0$. Already at accessibly low
temperatures, the string length may reach the sample width which is typically
$\sim1\mu m$.

\subsection{Combined topological defects and the NBN generation.}

A remarkable phenomena of sliding DWs is the generation of the so called
Narrow Band Noise (NBN) \cite{Gruner-CDW} which is a coherent periodic
unharmonic signal with the fundamental frequency $f$ being proportional to the
mean $dc$ sliding current $j$ (per chain) yielding the universal ratio of
$ef/j$. In CDWs ideally $ef/j=1/2$ which corresponds to carrying of two
electrons by displacing of the CDW by its wave length $\lambda$. In SDWs the
ratio has been accessed only indirectly \cite{Jerome:91,Clark:97} with different
experiments being in favor of either $ef/j=1/2$ or the twice higher ratio
$ef/j=1$. Surprisingly, there is no consensus on the origin of such a bright
and widely exploited effect as the NBN, with two main pictures
competing.
\newline i. The Phase Slip Generation (PSG) model suggests that the
NBN is generated by phase discontinuities of phase slip processes occurring
periodically near injecting junctions \cite{Ong:84,Ong:85,Gorkov:83,Gorkov:89,Gorkov:84,Batistic:84}
(and references to later experiments in \cite{Req}).
\newline ii. The earlier and better known
Wash-Board Frequency (WBF) model suggests that the NBN is generated
extrinsically in the course of the DW passing through the host lattice sites
or its defects \cite{Gruner-CDW,Zettle}. Recall that the oscillating densities
of the charge or of the lattice displacements in the CDW and the SDW are
different:
\[
\rho_{cdw}\sim\eta_{cdw}=A\cos[Qx+\varphi]\ \mathrm{but}\ \rho_{sdw}\sim
\delta(\vec{\eta}_{cdw})^{2}\Rightarrow A^{2}\cos[2Qx+2\varphi]
\]
- the charge modulation wave length in the SDW is only half of the CDW one
$\lambda=2\pi/Q$. Hence within the WBF model the NBN frequency is doubled for
the SDW.

The observation of the twice different ratios was considered to be in favor of
the WBF model which nevertheless has an unresolved weaknesses. The original
concept implied the interaction between the rigid DW and the regular host
lattice $\sim\cos(n\varphi)$ where the commensurability index is typically
$n=4$ which would give an $n$- fold WBF contrary to experiments. A later
belief was that the necessary potential $V_{imp}$ is provided by the host
impurities. But actually $V_{imp}\sim\cos(Qx_{i}+\varphi)$ so that the
positionally random phase shifts $-Qx_{i}$ will prevent any coherence in
the linear response.

An intermediate scenario suggests that the DW does not slide at the sample
side surface so that the coupling $\sim\cos(\varphi_{bulk}-\varphi_{surface})$
with $\varphi_{bulk}\propto t$ and $\varphi_{surface}=cnst$ would provide a
necessary WBF. This separation of moving and staying layers requires for a
sequence of dislocations thus building a bridge to a version of the PSG
scenario. The PSG model is attractive because the PSs are ultimately necessary already
to provide the current conversion at the contacts. A weak point of the PSG
model is to explain their regularity as shown by a remarkably high coherence
of the NBN in experiments.

\section{Two-fluid hydrodynamics for collective and normal variables in DWs.}

\subsection{Problems with Ginzburg -Landau like models and their time-dependent generalizations for density waves.}

Describing and modeling inhomogeneous states and transient processes in
DWs calls for an efficient phenomenological theory which should take
into account the degenerate order parameter, normal carriers and the electric
field. Related cases of vortices in superconductors and dislocations in
density waves \cite{Gorkov:83,Gorkov:89,Gorkov:84} are described usually
within Ginzburg-Landau (GL) like models and their time-dependent (TDGL)
generalizations. The essence of these approaches is that electrons are
excluded ("fermions are integrated out") at the earliest stage in favor of the
order parameter. While being consistent for the explicitly gauge-invariant
theory of superconductors \cite{Ivlev-Kopnin:84,Kopnin:01}, this approach
fails for density waves and other electronic crystals. The discrepancy is
particularly pronounced for processes with a dynamical vorticity - phase slips
or moving dislocations, leading to a non-physical production or elimination of
condensed particles in their cores. Working at all steps explicitly with
interfering order parameter and normal electrons, employing the chiral
invariance, and taking the account for the related quantum anomaly have allowed us
\cite{SB+NK:19-annals} to construct the treatable theory without descending to
the burdens of microscopic calculations. The price paid is that the theory is
non-analytical with respect to the order parameter which originates some
particular difficulties in numerical modeling. Nevertheless, with some
simplifications we succeed to demonstrate the dynamical creation of vortices
and phase slips processes, including pictures presented in this article.

In the TDGL theory for superconductors, the charge conservation is
automatically preserved because the equation of motion for the order parameter
contains only gauge invariant combinations of the phase and the
electromagnetic field. But in CDWs the order parameter phase characterizes
breaking of the chiral (translational) rather than the gauge symmetry. The
equations of motion for the CDW phase certify the local equilibrium of forces
(elastic, Coulomb, frictional) with no relation to the charge conservation.
Instead, the last is usually preserved automatically by construction of the
charge and the current densities:
\begin{equation}
n_{c}={\rho}_{c}{\partial}_{x}\varphi/\pi~,~j_{c}=
-{\rho}_{c}{\partial}_{t}\varphi/\pi
\label{EQ1}
\end{equation}
(to be compared with ${j}_{SC}\propto\rho_{sc}\partial_{x}\vartheta$ for the
current dependence on the phase ${\vartheta}$ in superconductors). (Here and
below the currents refer to the number of particles and the electronic charge $e$
is incorporated into the corresponding potentials.) The normalized (to $T=0$)
condensate density or the phase rigidity $\rho_{c}(A)$ has a
property that ${\rho_{c}(0)=1}$ and
$\rho_{c}(A)\sim{A}^{2}\rightarrow0$ at $A\rightarrow0$. The expressions
(\ref{EQ1}) can ensure the charge conservation automatically indeed, but
only if the CDW amplitude is invariable: $A(t,x)=cnst$, otherwise
\begin{equation}
\frac{dn_{c}}{dt}={\partial}_{t}n_{c}+{\partial}_{x}j_{c}=\frac{1}{\pi}
\left({{\partial}_{x}{\rho}_{c}\partial}_{t}\varphi-{{\partial}_{t}{\rho}_{c}\partial}_{x}\varphi\right)\neq0
\label{EQ2}
\end{equation}
The charge conservation ${dn_{c}/dt=0}$ is violated in the current-carrying
state ${\partial}_{t}\varphi\neq0$ if $\rho_{c}$ is not space homogeneous
{$\partial_{x}\rho_{c}\neq0$} and in the strained (charged) state
${\partial}_{x}\varphi\neq0$ if {$\rho_{c}$} varies in time {$\partial_{t}\rho_{c}\neq 0$}.
The effect is particularly disturbing in description of motion or
nucleation of vortices or of phase-slip processes when space and time
derivatives are high near the vortex core or the amplitude node. It looks that
for a general spacio-temporal regime there is no explicit way to define the
CDW collective charge and current via the order parameter alone. It is
ultimately necessary to take into account the normal~carriers explicitly,
without integrating them out prematurely. At first sight, that would require
for descending to the fully microscopic theory with its notorious
complications even in linear or gapless regimes \cite{Gorkov:89,Art}. Still,
there is a way \cite{SB+NK:19-annals} to keep the phenomenology which is based
on the knowledge of chiral invariance and transformation, importantly taking
into account the chiral anomaly.

The paradox is present also in theory of an arbitrary electronic crystal
beyond the quasi one-dimensionality and the particular microscopics inherent
to DWs.  E.g. for a Wigner crystal the charge and the current densities are
expressed via the field of displacements $\mathbf{u}(\mathbf{r},t)$
(which generalizes
$\mathbf{u\rightarrow}\varphi\mathbf{\hat{x}}$ of DWs) as
${n}_{c}=-\nu(\mathbf{\nabla}\cdot\mathbf{u})$,
$\mathbf{j}_{c}=\nu{\partial}_{t}{\mathbf{u}}$ where $\nu(\mathbf{r},t)$ is the filling
factor of the crystal unit cell by electrons. Then the conservation law is
violated for a variable $\nu(x,t)$:

\[
\frac{dn_{c}}{dt}={\partial}_{t}n_{c}+(\mathbf{\nabla}\cdot\mathbf{j}_{c})=
(\mathbf{\nabla}{\nu\cdot\partial}_{t}{\mathbf{u)}}{\mathbf{-}}
{\partial}_{t}\nu(\mathbf{\nabla}\cdot{\mathbf{u)}}\neq0
\]

\subsection{The tool of the chiral transformation and of the related anomaly.}

The Lagrangian $H-i\hbar{\partial}_{t}$ for electrons in 1D CDW has a form:
\begin{equation}
{\mathcal{L}}=\left(
\begin{array}
[c]{cc}
-i\hbar{\partial}_{t}-i\hbar v_{F}{\partial}_{x}+\Phi-v_{F}A_{x} & \Delta
e^{i\varphi}\\
\Delta e^{-i\varphi} & -i{\hbar\partial}_{t}+i\hbar v_{F}{\partial}_{x}
+\Phi+v_{F}A_{x}
\end{array}
\label{EQ6}
\right)
\end{equation}
where $\Phi$ and $A_{x}$ are the scalar and the vector potentials. The chiral
transformation ${\psi}_{\pm}\longrightarrow{\psi}_{\pm}{\exp}(\pm i\varphi/2)$
\cite{SB:76}, brings the wave function to the local frame of a distorted CDW
phase. It eliminates the phase factors in non-diagonal elements in
${\mathcal{L}}$, but in expense of additional parts in $\Phi$ and $A_{x}$:
\begin{align}
\Phi &  \longrightarrow V=\Phi{+}\hbar v_{F}/2{\partial}_{x}\varphi
~,~A_{x}\longrightarrow A_{x}^{\ast}=A_{x}+\hbar/(2v_{F}){\partial}_{t}\varphi\\
E_{x}  &  \longrightarrow E_{x}^{\ast}=-\partial_{x}V+\partial_{t}A_{x}^{\ast}=
E_{x}-\hbar v_{F}/2\left(\partial_{x}^{2}-\partial_{t}^{2}/v_{F}^{2}\right)\varphi
\label{EQ7}
\end{align}
The additions to $\Phi$ and $A_{x}$ from ${x,t}$ derivatives of the phase are
naturally interpreted as the Fermi energy shift $\delta E_{F}=v_{F}\delta
P_{F}$ following the Fermi momentum shift $\delta P_{F}=\hbar{\partial}_{x}\varphi/2$ under the CDW phase deformations. Under the CDW phase deformation and the applied electric field $E_{x}$, the electrons experience
the chiral invariant potentials ${V}$ and $A_{x}^{\ast}$ and the
longitudinal force $E_{x}^{\ast}$.

At first sight, we have arrived at the transparent picture of a 1D Dirac
semiconductor with the gap $2\Delta=2A\Delta_{0}$ under the effective
electric field (\ref{EQ7}), and it looks straightforward to exclude the
fermions to arrive at an effective action
 $S\{\Phi,\varphi,A\}=\int Wdxdt$.
And here we arrive at puzzling contradictions:

i. In view of Eq.(\ref{EQ7}), the free energy ${W}$ should contain the
potential and the phase only in the invariant combination ${V}$. Having choosen the phase as $(\hbar v_{F}/2){\partial}_{x}\varphi=-\Phi+$, the effective potential $V$ disappears from the
action, then no density and no polarization are perturbed with respect to
$\Phi$ which contradicts to basic properties of the CDW as both the
semiconductor with respect to electrons and the metal in the collective behavior.

ii. We definitely expect ${W}$ to contain the term $\propto\rho_{c}(\partial_{x}\varphi)^{2}$
where $\rho_{c}$ is a collective density
responsible for the phase rigidity. But treating the Lagrangian
(\ref{EQ6}) perturbatively with respect to the effective electric field
(\ref{EQ7}) we evidently should get
\begin{equation}
{\delta W}_{n}=-\frac{\varepsilon(k,\omega) }{8\pi e^{2}}{\left(E_{x}^{\ast}\right)}^{2}=
-\frac{\varepsilon(k,\omega}{8\pi e^{2}}
{\left(-\frac{\hbar v_{F}}{2}\frac{{\partial}^{2}\vartheta}{\partial x^{2}}+E_{x}\right)}^{2}
\label{EQ8}
\end{equation}
where $\epsilon$ is the dielectric
susceptibility as a function of the wave number ${k}$ and the frequency
$\omega$ (the last will be neglected here for shortness). At small ${k}$:
\begin{equation}
\epsilon=\epsilon_\Delta+\frac{1}{(lk)^2} \, , \,
\frac{1}{l^2}=\frac{\rho_n}{r_0^2}~,~\rho_n=\frac{dn}{d\zeta}N_F^{-1}
\label{EQ9}
 \end{equation}
Here $r_{0}$ is the Thomas-Fermi radius of the parent metal, ${n}$ is the
concentration of free electrons related to their chemical potential $\zeta$,
and $\rho_{n}$ is the normalized DOS which also will be found to be the normal
density. The dielectric constant $\epsilon_{\Delta}\propto({r_{0}\xi}_{{{0}}})^{-2}$ collects the polarization of electrons gapped by the CDW while
$\rho_{n}$ comes from conducting carriers thermally excited or injected
\cite{BKRM:00} above the gap providing the finite screening length
${l=r_{0}/\surd\rho}_{n}$. Both contributions to ${\delta W}_{n}$ bring
drastic contradictions: \\
a) The term with $\epsilon_{\Delta}$ in Eq.(\ref{EQ9}) yields to (\ref{EQ8}) the forth order gradients of the
phase ${({\partial}_{xx}^{2}\varphi)}^{2}$ instead of the expected second
order one $\propto{({\partial}_{x}\varphi)}^{2}$.\\
 b) The singularity in ${k}$ in the term with $\rho_{n}$ seems at first to serve fortunately by canceling
excess gradients leading to the contribution
$\propto{({\partial}_{x}\varphi)}^{2}$:\\
$\delta W_{\rho n}=-{\rho}_{n}({\hbar v_{F}/4\pi)\left({\partial}_{x}\varphi+\pi N_{F}\Phi\right)}^{2}$
which expression brings two confusions with respect to the expected
$W_{\rho c}={\rho}_{c}({\hbar v_{F}/4\pi)\left({\partial}_{x}\varphi\right)}^{2}$
- even the signs in expressions for $W_{\rho n}$ and $W_{\rho c}$ are opposite, and also
temperature dependencies are conflicting among coefficients $\rho_{n}$
(expected to rise from zero at ${T=0}$ up to $1$ at $T_{c}$) and {$\rho_{c}$}
(expected to fall to zero at ${T_{c}}$ starting from $1$ at ${T=0}$).

These contradictions can be traced back to the notion of the chiral anomaly
and they can be cured by properly taking this anomaly into account.
The chiral anomaly is ubiquitous to the premature linearization of electronic spectra.
First, in the course of the linearization, the control is lost of the position
of the bottom of the electronic band, hence of the difference between actions
of the external potential $\Phi$ and of the Fermi energy shifting $\delta E_{F}$
which wrongly appear to be additive. Moreover, the whole energy cost of
the chiral transformation (CT) ${\delta W_{CT}}$ is lost. This energy can be
captured from the non-linearized spectrum of the normal metal if we consider
the chiral transformation perturbation ${\delta n_{CT}=\partial}_{x}\varphi/\pi$
as a redistribution of the total particle density accompanying the this transformation.
\begin{equation}
\delta W_{CT}=\frac{\hbar v_{F}}{4\pi}{\left({\partial}_{x}\varphi\right)}^{2}
+\frac{\Phi}{\pi}{\partial}_{x}\varphi
\label{EQ13}
\end{equation}
where the first term is the density perturbation cost $({\delta n_{CT})^{2}/(2N_{F})}$
and the second term is its potential energy $\delta n_{c}\Phi$.
Beyond these physical arguments \cite{SB:93-1,SB:93-2}, the derivation
of the chiral anomaly in the spirit of the field-theory procedure of
regularization of fermionic determinants was demonstrated for CDWs
\cite{Krive:87} and for SDWs \cite{Krive:87,Dupuis:00}, recall also the special field-induced SDWs \cite{Yak:98}. In the Appendix we shall briefly clarify the microscopic origin of the chiral anomaly paradox.

Bringing together the nonperturbative contribution (\ref{EQ13}) and the perturbative one (\ref{EQ8}), we get
\begin{equation}
W_{tot}={\delta W}_{\rho n}+\delta W_{CT}=
{\rho}_{c}\left(\frac{\hbar v_{F}}{4\pi}{\left({\partial}_{x}\varphi\right)}^{2}+
\frac{\Phi}{\pi}{\partial}_{x}\varphi\right)  -
{\rho}_{n}\frac{{\left(\Phi\right)}^{2}}{\pi\hbar v_{F}}
\label{EQ14}
\end{equation}
where ${\rho}_{c}=1-{\rho}_{n}$. The above equation correctly manifests the expected dependencies in $T$ and $A$ yielding
also the important relation $\rho_{c}+\rho_{n}=1$.

The total charge density becomes
\begin{equation}
n_{tot}=\frac{{\partial W}_{tot}}{\partial\Phi}=
\frac{1}{\pi}{{\rho}_{c}\partial}_{x}\varphi-{\rho}_{n}\Phi N_{F}=
\frac{1}{\pi}{\partial}_{x}\varphi-{\rho}_{n}VN_{F}
\label{n_tot}
\end{equation}
Here the first form of $n_{tot}$ interpret the CDW charge as ${\rho}_{c}$
leaving the electronic density to react only to $\Phi$ which approach is
common and convenient while ambiguous: with a variable ${\rho}_{c}$ this form
leads to violation of the charge conservation as it was discussed in the
previous section. The second form of $n_{tot}$ lets us to understand that the
collective charge density is always the nominal one embracing all electrons in
the vacuum and excited states, not reduced by the factor {$\rho_{c}$};
meanwhile the charge density of normal electrons is perturbed by their actual
combined potential ${V}$ rather than by its part {$\Phi$}. In this picture,
the charge and the current are given by independently conserved counterparts
as show in Eq. (\ref{EQ1}).

The above illustrative discussion was valid in lowest bilinear approximation in gradients and potentials and, more restrictively, for the constant amplitude $A$, hence $\rho_n,\rho_c$. In the next section we shall suggest a general nonlinear scheme necessary for modeling configurations with vortices.

\subsection{General equations for nonlinear regimes.}

The above relations written for the limit $A=cnst$ and for small deviations of
${n_{e},n_{h}}$ can be generalized by extending the energy functional as $\int
Wdx$ with
\begin{align}
W=\frac{\hbar v_{F}}{4\pi}\left[{\kappa}_{x}{\left({\partial}_{x}A\right)}^{2}+
{\kappa}_{\perp}{\left(  {\nabla_{\perp}}A\right)  }^{2}+
{\kappa}_{\perp}A^{2}{\left({\nabla_{\perp}}\varphi\right)}^{2}\right]+
\label{EQ15}    \\
\left\{\frac{\hbar v_{F}}{4\pi}{\left({\partial}_{x}\varphi\right)}^{2} +
\frac{1}{\pi}\Phi{\partial}_{x}\varphi\right\}  +\left(\Phi{+}
\frac{\hbar v_{F}}{2}{\partial}_{x}\varphi\right)  n+{F}(A,n_{e},n_{h})-
\frac{{\varepsilon}_{host}s}{8\pi}{\left(\nabla\Phi\right)}^{2}
\nonumber
\end{align}
Here the parameter $\kappa_{\perp}$ is the CDW share modulus coming from the
interchain coupling of CDWs, and the on-chain rigidity of the amplitude
$\kappa_{x}\sim1$ (to be put ${\kappa}_{x}=1$). $F(A,n_{e},n_{h})$ is
a free energy as a function of the normalized gap value
$A=\Delta/\Delta_{{0}}$ and of the concentration of normal carriers: electrons ${n_{e}}$ and holes
${n_{h}}$ with ${n=n_{e}-n_{h}}$. The equilibrium value $A_{eq}$ is
connected with ${n_{e},n_{h}}$ via the minimum of $F(A,n_{e},n_{h})$, in such
a way that$A_{eq}(n_{e},n_{h})$ vanishes when $n_{e,h}$, or better
say their chemical potentials $\pm\zeta$ surpass critical values, hence the
metallic phase with $A=0$ is restored. The terms in brackets $\{\}$ are
originated by the chiral anomaly of Eq. (\ref{EQ13}) coming from
background deformations of the CDW phase, the next term $\sim n$ comes from the
energy of intrinsic electrons in the combined potential $V$.

Assuming the dissipative regime for both $\varphi$ and $A$, functional
derivatives of (\ref{EQ15}) yield equations for the time evolution and
the Poisson equation for $\Phi$:
\begin{align}
{\partial}_{x}^{2}A+{\kappa}_{\perp}{\nabla_{\perp}^{2}}A+{\kappa}_{\perp}A{\left({\nabla_{\perp}}\varphi\right)}^{2}-\partial F/\partial A
={\gamma}_{A}{\partial}_{t}A
\label{EQ16}
\\
{\partial}_{x}^{2}\varphi+\pi N_{F}{\partial}_{x}\Phi+
\pi{\partial}_{x}n+{{\kappa_{\perp}}\partial}_{y}
\left(A^{2}{\nabla}_{\perp}\varphi\right)= {\gamma}_{\varphi}{\partial}_{t}
\label{EQ17}
\\
{\partial}_{x}\varphi/\pi+n =-{N_{F}r_{0}^{2}\nabla}^{2}\Phi
\label{EQ18}
\end{align}
Here $\gamma_{\varphi}=\gamma A^{2}$, $\gamma=cnst$, $\gamma_{A}=cnst$ are the
damping coefficients; $\gamma_{\varphi}$ is related to the sliding CDW
conductivity $\sigma_{CDW}$ as $\gamma_{\varphi}\sigma_{CDW}=N_{F}e^{2}/s=1/(4\pi r_{0}^{2}$).

In spite of a superficial similarity, these eqs.
show striking differences with respect to commonly use TDGL ones. Thus, in the
first Eq. the conventional term ${\kappa}_{\perp}A{\left(\nabla_{\perp}\varphi\right)}^{2}$
does not find its partner with the $x$ derivative of
$\varphi$ as if the longitudinal gradient of $\varphi$ does not suppress the
amplitude - the phase-slip nodes would not appear then. In the second and the
third eqs. the terms in brackets $\{\}$ {containing ${\partial}_{x}\varphi$,
${\partial}_{xx}\varphi$, and ${({\partial}_{x}\varphi)}^{2}$
are not multiplied by $A^2$, so the attempt to present them as
derivatives of $\Psi$ will bring $\sim1/A$,$\sim 1/A^{2}$ singularities:
contrary to conventional GL-type equations, now Eqs.
(\ref{EQ16},\ref{EQ17},\ref{EQ18}) are nonanalytic in the order parameter $\eta=$ $A\exp(i\varphi)$ or in
other words singular in its amplitude. Nevertheless, there are hidden
cancellations allowing to compensate for singularities, even if implicitly,
which is vitally important for allowance of space- and spacio-temporal
vortices. With additional approximations, below these compensations will be
better exposed or proved at least.

Finally, the kinetics of normal carriers can be taken in the
quasi-equilibrium diffusive approximation:
\[
\nabla\left(\hat{\sigma}_{n}\nabla\mu\right)=(e^{2}/s){\partial}_{t}n \, , \,
\mu_{n}=\zeta+\Phi+{\partial}_{x}\varphi/(\pi N_{F})
\]
 with the electrochemical potential
 $\mu_{n}$ and the conductivity tensor $\hat{\sigma}_{n}$.

For the instantaneous reaction of normal carriers in the limit  $\sigma_n\rightarrow\infty$ requesting for $\mu_n=0$, this relation helps to understand the decomposition of the total charge density ${n}_{tot}$. Differentiating over $x$ the general expression for ${n}_{tot}$ as given in the LHS of Eq.(\ref{EQ18}) and using $\mu_n=0$ we arrive at
\[
\partial_{x}n_{tot}=-\rho_{n}N_{F}{\partial_{x}}\Phi+\rho_{c}/\pi{\partial}_{x}^{2}\varphi
\]
For ${\rho_{n},\rho_c=cnst}$ this expession can be integracted to the restricted form of Eq.(\ref{n_tot}).
 Here the first term corresponds to the conventional reaction (the screening of the

\subsection{The limits of the local electroneutrality together with the
infinite conductivity.}
The general equations can be substantially simplified and clarified in two
limits: the infinite conductivity of normal carriers which yields
$\mu_{n}=\zeta+\Phi+{\partial}_{x}\varphi/(\pi N_{F})=0$ and the local
electroneutrality ${r_{0}\rightarrow0}$ which yields ${\partial}_{x}\varphi+\pi n=0$.
Separately the two limits have been considered in
\cite{SB+NK:19-annals}; here we shall remind only their joint result. The
above local relations allow to express ${\Phi,\partial}_{x}\varphi$ via
$\zeta$ (or equivalently via $n$) alone: $\Phi=n/N_{F}-\zeta$,$~{\partial}_{x}\varphi=-\pi n({\zeta,A)}$.
Then we arrive at two equivalent forms of equations for the phase:
\begin{equation}
{\partial}_{x}{\Phi}+(\pi{N_{F})}^{-1}\nabla_{\perp}A^{2}\nabla_{\perp}\varphi
-\gamma_{\varphi}{\partial}_{t}\varphi   =0~,~\Phi={(n/N_{F}-\zeta})
\label{EQ30}
\end{equation}

\begin{equation}
\frac{\rho_{c}}{\rho_{n}}{\partial}_{x}^{2}\varphi+{\kappa}_{\perp}
{\nabla_{\perp}}\left(  A^{2}{\nabla_{\perp}}\varphi\right)
 -{\gamma}_{\varphi}{\partial}_{t}\varphi
=\pi N_{F}\frac{\partial\zeta}{\partial A}{\partial}_{x}A \label{EQ31}
  \end{equation}

The behavior of different terms in Eq. (\ref{EQ30}) at small $A(t,x)$ is consistent since at
$A\rightarrow 0$ the combination
${n-\zeta N_{F}=\Phi N_{F}\rightarrow0}$ by construction and ${\gamma}_{\varphi}\rightarrow0$ by definition.
Curiously, the Eq. (\ref{EQ30}) does not show explicitly the commonly assumed longitudinal phase rigidity
${\propto\partial}_{x}^{2}\varphi$; it is hidden in the first term
$\partial_{x}\left(\pi n-\zeta\right)  $ implicitly, via the relation of
${\partial}_{x}\varphi+\pi n=0$. This term provides also the driving force for
the CDW current which can be written as $-\rho_{c}\partial_{x}\zeta$ instead
of the conventionally supposed $\rho_{c}E$: gradient of the normal carriers
chemical potential rather than the electric field.

The second form (\ref{EQ31}) shows explicitly the physical phase rigidity
accumulating all effects of normal carriers and Coulomb interactions. In the
first term in the LHS {the ratio {$\rho_{c}/\rho_{n}$} controls vanishing of
the phase rigidity at $A\rightarrow$}${0}$ approaching the metallic state and
the Coulomb hardening \cite{Art,Pouget:92} of the charged phase deformations
which dramatically increases with freezing out of screening by normal carriers
when $\rho_{n}\rightarrow0$. Remarkably, the form (\ref{EQ31}) does not
show any driving force for $\varphi$, apart from the less important term in
the RHS coming from the gradient of the amplitude $A$. The drive will come
only from the boundary conditions for the electric potential transferred to
the phase via the local relations of $\partial_{x}\varphi$ and $\Phi$ mediated
by $n$.

In the 1D regime we can exclude the phase by differentiating the Eq.
(\ref{EQ30}) over ${x}$ to arrive at the closed eq. for the pair of only
$n$ and $A$ variables defining the carriers' free energy $F$:
\begin{align*}
\gamma{\partial}_{t}n  &  ={{\partial}_{x}(\ A^{-2}{\partial}_{x}(N}_{F}\partial F/\partial n-n))\\
{\gamma}_{A}{\partial}_{t}A  &  ={\partial}_{x}^{2}A-\partial F/\partial A~,~F=F(n,A)
\end{align*}

Imposing only the limit of the electroneutrality while keeping $\sigma_{n}$ to
be finite, the driving field for the normal current becomes
$-\nabla\mu_{n}\rightarrow-\nabla(\Phi+\zeta-n/N_{F})=
${$-\nabla\Phi-(\rho_{c}/\rho_{n})\nabla n$}, i.e. the effective diffusion coefficient is enhanced as
$\rho_{c}/\rho_{n}$ which result could hardly be expected intuitively.
In the 1D regime the diffusion equation and the equation for the phase can be
combined and integrated once to yield the equation for the phase alone
\begin{equation}
\frac{1}{\pi}\left(\frac{1}{{\sigma}_{CDW}}+\frac{1}{{\sigma}_{n}}\right)
{\partial}_{t}\varphi{+{\partial}_{x}(\frac{1}{\pi}\partial}_{x}\varphi+\zeta)
=\frac{-1}{{\sigma}_{n}}J(t)
\label{EQ28}
\end{equation}
with $\zeta=\zeta(n,A)$ at $n=-\partial_{x}\varphi/\pi$.
The RHS of Eq. (\ref{EQ28}) gives the driving force as the total current
${J(t)=j}_{c}+j_{n}$ while multiplied by the normal resistance alone. The
collective and the normal resistivities combine additively to the effective
friction coefficient - the resistivity - of the moving CDW.

\subsubsection{Numerical modeling.}

There are some technical challenges in numerical implementations of the quoted
above equations which one commonly does not meet working within conventional GL
approaches, see e.g. \cite{Batistic:84, Kopnin:88}. The first is the control of
compensations at $A\rightarrow{0}$ in expressions for total charges,
currents, and the condensate energy bringing to action the hidden function of
the condensate density $\rho_{c}$. The second is the entanglement in dependencies
of thermodynamic functions and their derivatives on $A$ and on $n$ or $\zeta$:
approaching of $n$ or $\zeta$ to critical values should eliminate the
energy minimum over $A$ at $A\neq0$ opening the metallic state, e.g. in the
vortex core.

The best, and may be the only analytically transparent, advancing is possible
with the simplest Landau type expression for $F$:

\begin{equation}
F(n,A)=n^{2}/(2N_{F})+(-\tau+(n/n_{cr})^{2})(A\Delta_{0})^{2}N_{F}/2+bA^{4}\Delta_{0}^{2}N_{F}/4 \label{EQ33}
\end{equation}
where $a,b\sim1$. In the Eq. (\ref{EQ33}) the first term is the contribution of the normal metal, while two other
terms give the Landau type expansion in the order parameter with the proximity
$\tau\propto(1-T/T_{c})$ to the transition temperature being shifted by
presence of the normal carriers which critical concentration is ${n_{cr}}$.

We have performed a numerical modeling employing the following combination of equations: Eq.(\ref{EQ15}) for $A$ with  $F$ given by Eq.(\ref{EQ33}), Eq.(\ref{EQ16}) for $\varphi$ and the dissipative equation for $\vec{m}$ generated by the functional (\ref{W-spin}). The vortices can be spontaneously generated only if equations are written in invariarian variables, rather than in the economical form for the non-unique phase $\varphi$ and angle $\theta$.
The order parameter $\vec{\eta}_{sdw}=A\vec{m}\exp(i\varphi)$ is written as $\vec{\eta}_{sdw}=(u+iv)\{p,q\}\}$. For the CDW or for frozen alligned spins in SDW the invariant variables are just $u,v$. For a SDW with a spin vorticity the allowed varibles were taken as a set of four bilinear combinations $\{\alpha,\beta,\gamma,\delta\}=\{up,uq,vp,vq\}\}$ imposing the apparent constaint $\alpha\delta=\beta\gamma$. Examples of resulting calculations are shown in Figs.1,2,4a,b.

\section{ Conclusions }

A necessity of semi-vortices in conventional antiferromagnets in presence of
frozen-in host lattice dislocations was understood already by IED in
\cite{IED:77}. In the SDW the semi-vortices become the objects of the lowest
energy created in the course of phase slip process; the normal dislocation must split
into two objects of the combined topology with the repulsion between them. The
combined topological objects where the spin rotations are coupled to DW
displacements are stabilized  by lowering the Coulomb energy of dislocations. This combination
effectively reduces the SDW period allowing e.g. for the twice increase in the
NBN frequency, which is an important disputable question.

In presence of spin anisotropy the free rotation of spins is prohibited at
large distances from vortices, then the two objects are connected by a string
which is the Neel domain wall. Because of the high Coulomb energy of the phase
component, the regime of the confinement due to the domain wall is pushed away
beyond the distance of the screening length provided by the normal carriers
which can be arbitrarily large at low $T$. This allow the combined
semi-vortices to be unbound in realistically thin samples.

Exploiting the chiral transformation and understanding the role of the chiral
anomaly allows to formulate a phenomenological theory in terms of equations
for the DW complex order parameter, the electric potential and the
concentration of normal carriers. This approach resolves the problem of
violation of the conservation law for condensed carriers rising dangerously
for nonstationary inhomogeneous regimes; that allows to model consistently
such strongly nonlinear effects as phase slips and nucleation and propagation
of phase vortices. Conceptually, the unconventional view is that the
collective density and the current always correspond to the nominal number of
condensed electrons, as if the temperature is zero and there are no excitations
above the gap. The actually present normal carriers are dragged by the phase
deformations in such a way that their reaction erases the collective
quantities.

\section{Appendices}
\subsection{Appendix A. Anomalous energy and shape of the charged dislocation in a DW.}

In the Fourier representation, the energy of phase deformations looks like
\begin{equation}
W\{\varphi\}=\frac{\hbar v_{F}}{4\pi}\sum|\varphi_{k}|^{2}\left[
C_{\parallel}^{0}k_{\parallel}^{2}+C_{\bot}k_{\bot}^{2}+
\frac{r_{0}^{-2}k_{\parallel}^{2}}{(k_{\parallel}^{2}+k_{\bot}^{2}+r_{scr}^{-2})}\right]
\label{W-F}
\end{equation}
where $r_{0}=\sqrt{\hbar v_{F}s\epsilon/(8e^{2})}$ is the Thomas-Fermi
screening length in the parent metal which is the atomic-scale length, and
$r_{scr}=r_{0}/\sqrt{\rho_{n}}$ is the actual screening length by remnant
charge carriers in the DW state. Here the normalized normal density
$\rho_{n}\rightarrow1$ at $T\rightarrow T_{c}^{0}$ and $\rho_{n}\sim\exp(-\Delta/T)$
at low $T$ being activated across the DW gap $2\Delta$.
$\rho_{c}=1-\rho_{n}=C_{\parallel}^{0}$ is the condensate density.
Because of the very large Coulomb energy,
the longitudinal gradients are strongly suppressed and we can neglect in (\ref{W-F})
all appearances of $k_{\parallel}$ except for the combination $r_{0}^{-2}k_{\parallel}^{2}$.
At large transverse distances beyond the screening length $r_{\perp}\gg r_{scr}$
we can neglect also $k_{\bot}^{2}$ in the denominator in (\ref{W-F}); then the effective elastic theory is restored,
while in stretched coordinates $(x\sqrt{C_{\bot}\rho_{n}/\rho_{c}},r_{\perp})$:
\begin{equation}
W\{\varphi\}\approx\frac{\hbar v_{F}}{4\pi}\sum|\varphi_{k}|^{2}
\left[C_{\parallel}k_{\parallel}^{2}+C_{\bot}k_{\bot}^{2}\right]  \, ,\; C_{\parallel}=\rho_{c}/\rho_{n}\,,\;k_{\bot}\ll r_{scr}
\end{equation}

Within the screening distance $r<r_{scr}$, the expression (\ref{W-F}) describes a nonanalytic elastic theory,
already known for uniaxial ferroelectrics, with the energy dependent on ratio of gradients rather on
their values. In this nonscreened regime the energy dependence on the
transverse length $L_{\bot}=k_{\bot}^{-1}$ at any given longitudinal scale
$L_{\Vert}=k_{\Vert}^{-1}$ is not monotonous, showing a minimum at
$L_{\bot}^{2}\sim L_{\Vert}r_{0}C_{\bot}^{1/2}$. Multiplying the minimal energy
density $\sim C_{\bot}L_{\bot}^{-2}$ by the characteristic volume $L_{\Vert}L_{\bot}^{2}/a_{\perp}^{2}$
we estimate the energy as a function of $R=L_{\bot}$ as
\[
C_{\bot}^{1/2}R^{2}\hbar v_{F}/r_{0}=w_{C}N
~\mathrm{with}~w_{C}\sim T_{c}a_{\perp}/r_{0}
\, \mathrm{at} ~ N=\pi R^{2}/a_{\perp}^{2}
\]

Consider a D-loop of a radius $R$ in the $(y,z)$ plane embracing a number
$N=\pi R^{2}/a_{\perp}^{2}$ chains or a D-line stretched in $z$ direction at a
distance $Y=a_{\bot}N$ from its counterpart or from the surface. For the DL
energy $W_{D}(N)$ the above estimations, with precise values found in
\cite{SB+SM-disl:92}, can be summarized as
\newline\noindent  1. $W_{D}(N)\sim T_{c}\sqrt{N}\ln(N)$, valid within length $r_{0}$ - actually only at $N\sim1$ i.e.
for a elementary loop embracing only one chain, known as $2\pi$ soliton.
A similar expression is valid at all distances for the spin vortex.
\newline\noindent  2.
$W_{D}(N)\sim\ Nw_{C}$, $w_{C}\sim T_{c}a_{\bot}/r_{0}$, valid at
$r_{0}<R<r_{scr}$.
\newline\noindent 3. $W_{D}(N)\sim T_{c}(r_{scr}/r_{0})\sqrt{N}\ln(Na_{\perp}/r_{scr})$ at $R>r_{scr}$.
\newline Within the wide (at low $T$) regime 2. the dislocation energy follows the confinement-type area law W$\sim N$
rather than the usual perimetrical-logarithmic law $\sim\sqrt{N}\ln N$.

\subsection{Appendix B. The microscopic origin of the chiral anomaly.}

Technically (see \cite{Krive:87}), the anomaly appears from uncertainty in
calculations of fermionic determinants needing to be regularized, which
procedure is usually treated in a relativistically invariant way by traditions
of the field theory. It is instructive to see the origin of the anomaly in a
transparent way related to rules of many-electronic systems, namely to the
Tomas-Fermi procedure. Let the electrons in the parent metal occupy a
parabolic spectrum ${p^{2}/2m}$ up to a Fermi energy $E_{F}$.
In a smooth potential ${V(x)}$ the electronic wave functions $\Psi_{E}$
and their density distributions for an eigen energy ${E}$ are legitimately
given by the WKB expressions
\[
\Psi_{E}\propto(E-V(x)^{-1/4}\exp(i\int dx^{\prime}(E-V(x^{\prime1/2})
~,~\rho_{E}=|\Psi(x)|^{2}=(E-V(x)^{-1/2}
\]
Summing up ${\rho}_{E}$, the total density can be obtained and expanded in
$V/E_{F}$ as $\rho(x)-<\rho>\approx -V(x)N_{F}$ which yields the most
important characterization of a metal: the linear response of the density to
the applied potential. But if the spectrum linearization is done prematurely, then the
Schr\"{o}dinger eq. becomes the first order one, the wave function $\Psi_{E}$
looses the prefactor, ${\rho(x)}$ becomes just constant, and the reaction to
the local potential disappears. The role of the chiral anomaly action (${W}_{CA}$ in Eq. (\ref{EQ13}) is just to restore this missing contribution.

\end{document}